\documentclass[aps,twocolumn]{revtex4}
\usepackage{epsfig}

\newcommand{\beq}{\begin{equation}}
\newcommand{\eeq}{\end{equation}}
\newcommand{\bea}{\begin{eqnarray}}
\newcommand{\eea}{\end{eqnarray}}
\begin{document}
\draft
\title{Scale Free Cluster Distributions from Conserving Merging-Fragmentation Processes.}
\author{Jesper Ferkinghoff-Borg$^{\star,\dagger}$, Mogens H. Jensen$^\dagger$, Joachim Mathiesen$^{\dagger,\ddagger}$ and
Poul Olesen$^\dag$,\footnote{email: borg,mhjensen,mathies,polesen@nbi.dk} }
\affiliation{$^\star$NORDITA, Blegdamsvej 17, DK-2100 Copenhagen {\O},
  Denmark \\
$^\dagger$Niels Bohr Institute,
Blegdamsvej 17, DK-2100 Copenhagen {\O}, Denmark \\
$^\ddagger$NTNU, Institute for physics, N-7591 Trondheim, Norway}
\date{\today}
\begin{abstract}
We propose a dynamical scheme for the combined processes of fragmentation 
and merging as a model system for cluster dynamics
in nature and society displaying scale invariant properties. 
The clusters merge and fragment with rates proportional to 
their sizes, conserving the total mass. The total number of clusters
grows continuously but the full time-dependent distribution can
be rescaled over at least 15 decades onto a universal curve 
which we derive analytically. This curve includes a scale free
solution with a scaling exponent of $-3/2$ for the cluster sizes. 
\end{abstract}

\maketitle

\vskip0.3cm

The combination of the two basic dynamical processes, fragmentation
and merging, is of importance in a number of situations of nature
and society. Examples range from the dynamical evolution of companies
where mergers and split-offs are important ingredients in daily 
business life. Empirical observations of US companies indicates that
their sizes follow a scale invariant distribution \cite{company}. 
Another example comes from the dynamics of grain boundaries in crystal
growth where neighboring grains might merge at the same time as
other grains fragment away from larger crystals. In crystal growth
diffusion of grain boundaries can also be of importance in an interplay
with merging and fragmentation \cite{is,ferkinghoff,poul}. Fish schools are
known to merge together resulting in scale free distribution
up to a school size after which it becomes
exponentially distributed \cite{bonabeau}. In polymer
physics merging and fragmentation are of importance in relation
to gel and shattering transitions \cite{ziff}; other examples are
aerosols \cite{frielander} and scaler transport \cite{melzak} 
fragmentation/coagulation (i.e. merging) models
(see e.g. \cite{drake,leyvraz}
for reviews on merging/fragmentation processes).
Finally, we mention an astrophysical application for the formation
of stars and interstellar clouds \cite{source}.

It is known empirically that the resulting cluster
distribution in some of the above mentioned examples has 
scale free behavior. There has however, to the best of our knowledge,
neither been proposed an analytical theory nor any fundamental model studies 
of these scale free distributions in closed systems driven by
combined merging/fragmentation processes.
In this Letter we respond to this challenge by introducing a dynamical
model system where clusters merge and fragment with rates proportional
to the sizes, conserving the total mass of the system. Indeed, some previous models
have been introduced to explain the scale free behavior but they
relay on a source term for small cluster sizes and thus represent
systems with non-conserved mass \cite{source,vicsek,family,ernst,zheng,minn}. 
The dynamics of our model self-organizes the system to possess 
a stationary ``continuous" scale
free cluster size distribution over many decades and up to a
characteristic size above which only the largest clusters will be
situated. We introduce an analytical theory for the model based on Smoluchowski's 
continuous cluster fragmentation and coagulation (i.e. merging) model 
\cite{smoluchowski1916}. We show that this model with ``mass" kernels exhibits
an exact scale invariant solution of the cluster distribution,
$n(x) \sim x^{-3/2}$, at a critical point where merging processes balance
the fragmentation processes. In this solution, it turns out
that there is a gap in the distribution above which only a few large clusters exist.
The scaling exponent $-\frac{3}{2}$ is in agreement with distributions
of fish schools \cite{bonabeau} and many critical branching processes
\cite{branching}, for instance.
 
Our model is defined in terms of ${\cal N}$ clusters $\{\tilde x_i\}_{i=1}^{{\cal
    N}}$ each of size $\tilde x_i$, where tilde ($\tilde{~}$) in the following 
is used for dimensionfull quantities. The total initial mass of all clusters is thus 
$\tilde M = \sum_{i=1}^{\cal N} \tilde x_i$. At time $\tilde t$ the clusters
$i$ and $j\neq i$ are chosen to merge
with a rate  $K_{ij}=\beta \tilde x_i \tilde x_j$ and the cluster $k$
is chosen to fragment with the rate $F_k=f \tilde x_k $. The size 
of the two clusters resulting from a fragmentation is chosen randomly
    (ie. with a uniform distribution).
The average time $\delta \tilde t$ for the next merging or fragmentation
event to occur is obtained from the total merging and
fragmentation rates, $\tilde k_{\mbox{merge}}$ and $\tilde k_{\mbox{frag}}$,
as  $\delta \tilde t=(\tilde k_{\mbox{frag}}+\tilde k_{\mbox{merge}})^{-1}$,
where 
$$
\tilde k_{\mbox{merge}}=\sum_i \sum_{j>i} K_{ij} =
\beta/2\left( \tilde{M}^2 - \tilde M_{2} \right).
$$
and 
$$
\tilde k_{\mbox{frag}}=\sum_i F_i = f\tilde{M} 
$$
and $\tilde M_2=\sum_i \tilde x_i^2$ is the second moment of the
distribution. In time-step $\delta \tilde t$ the merging and fragmentation 
processes are thus associated with the probabilities
$p_{ij}=K_{ij}\delta \tilde t$ and $p_k=F_k\delta \tilde t$, respectively.
Since the processes preserve the total mass we can
express the dynamics in terms of the dimensionless quantities,
\begin{equation}
\begin{array}{rlrl}
x &=\frac{\tilde x}{\tilde M}, & t &=\tilde t(f\tilde{M})^{-1} \\
k_{\mbox{frag}} &=1, & k_{\mbox{merge}} &=b/2(1-M_2), \\
\end{array}
\label{dimless}
\end{equation}
where $b=\tilde M\beta/f$ and $M_2=\sum_i x_i^2$.

\begin{figure}
\epsfig{file=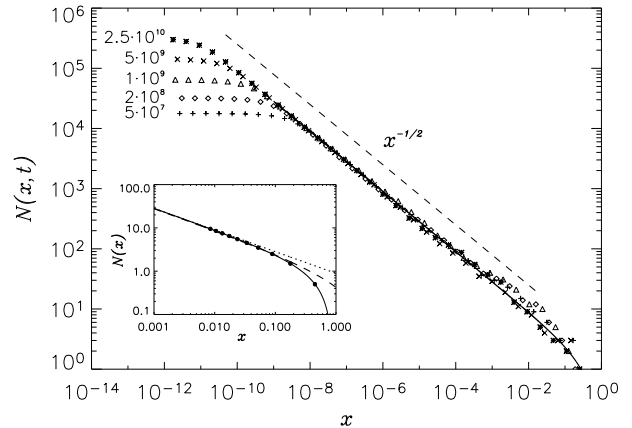,width=.5\textwidth}
\caption{ 
Cumulative distribution, $N(x,t)$, of the fragment sizes $x$ at 
various times, $t=5\cdot 10^7, 2\cdot 10^8, 10^9,5\cdot 10^9,2.5\cdot
10^{10}$. The thick solid line is the average cumulative distribution,
$N(x)$, for $x>10^{-9}$ taken over times $t>10^9$. At any instant,
$N(x,t)$ displays a characteristic cross-over from an
almost flat distribution to a scale-invariant distribution,  $N(x)\sim x^{-1/2}$,
at larger $x$. As time goes the scale free part of the distribution extends to smaller
and smaller scales. The inset shows the behavior of $N(x)$ in 
the large part of the mass spectrum. The dotted line is the
simple scaling form, $N_1(x)=Ax^{-1/2}$, with $A\approx 0.9$, 
and the dashed line is the predicted cumulative
distribution, $N_0(x)=A\left( x^{-1/2} -3/16 A\pi \right)$ (see text).
The dots in the figure represent the typical size ($50\%$ fractile) of
the 10 largest fragments, ie. points corresponding to $N(m_i)=i-1/2$, where $m_i$ the
the typical size of the $i$'th largest fragment. The predicted
cumulative distribution, $N_0$, deviates from data around $x=l\sim
0.1$ corresponding to the typical size of the third largest fragment,
$m_3\approx 0.09$. The naive scaling 
form, $N_1$, deviates from data already around $x\sim 0.02$.} \label{snapshot}
\end{figure}

\begin{figure}
\epsfig{file=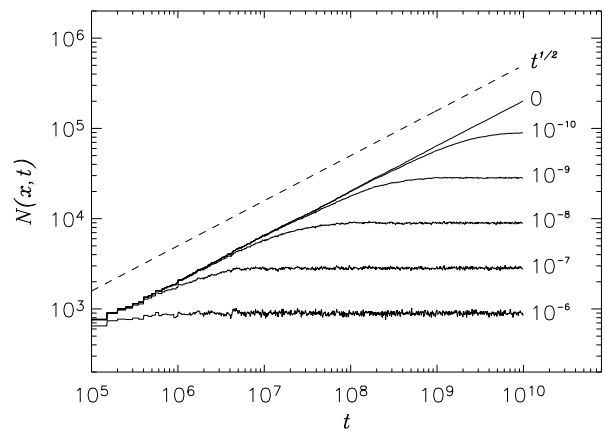,width=.5\textwidth}
\caption{
Cumulative distribution, $N(x,t)$, as function of 
time $t$ at various sizes, $x=0,10^{-10},10^{-9},10^{-8},10^{-7},10^{-6}$. 
For each positive value of $x$, $N(x,t)$ displays a cross-over from 
a power law $N(x,t)\sim t^{1/2}$ at short times
to a constant, $N(x,t)=N(x)$, at large times.}\label{evolution}
\end{figure}

In Fig. \ref{snapshot} we show the cumulative distribution,
$N(x,t)=\sum_i \Theta(x_i(t)-x)$, for a simulation with 
$b=3$ at various times $t$. Here, $\Theta(x)$ is the Heaviside function, $\Theta(x)=1$ for $x\ge
0$ and $\Theta(x)=0$ for $x<0$. At each time $t$, the fragments
larger than a characteristic cross-over scale, $x\gg x_c(t)$, are 
distributed according to a power law, $N(x)\sim x^{\alpha}$
with $\alpha=-1/2$. As seen in the figure, $x_c(t)\rightarrow 0$ for $t\rightarrow
\infty$, implying the existence of a scale invariant state which
extends to smaller and smaller scales as time increases. 
The same data are shown in Fig. \ref{evolution} with $N(x,t)$  as
function of $t$ for various values of $x$. Interestingly, the total number of domains ${\cal N}(t)=N(0,t)$ grows 
as ${\cal N}(t)\sim t^\gamma$ with the power $\gamma=-\alpha=1/2$. For each
$x>0$, the distribution is time-dependent, $N(x,t)\sim t^{1/2}$, for
times smaller than a characteristic cross-over time $t\ll t_c(x)$, and 
becomes time independent, $N(x,t)=N(x)$ for $t\gg t_c(x)$.

Simulations for other values of $b>2$ lead to distributions, $N(x)$, and evolutions ${\cal N}(t)$ 
which are very close to power-laws, $N(x)\sim x^{\alpha(b)}$ and ${\cal N}(t)\sim
t^{\gamma(b)}$ with $b$-dependent exponents $\gamma(b)$ and $\alpha(b)$ satisfying $\gamma(b)=-\alpha(b)$
\cite{usnew}. Nonetheless, only at $b=3$ does the scale-invariant
behavior become exact. For $b\le 2$, the maximum mass $m_1$ 
approaches $m_1=0$ as $t\rightarrow \infty$ \cite{usnew}, 
implying that no time-independent solution exists in this parameter
region. 

To analyze the origin of the scale invariant solution we
express the dynamics in terms of Smoluchowski's continuous cluster 
fragmentation and coagulation (i.e. merging) model \cite{smoluchowski1916}.
We have previously presented this type
of rate equations including diffusion, fragmentation and coagulation
(ie. merging) terms, which are of relevance for dynamical processes
of ice crystals, alpha helices and macromolecules
\cite{is,ferkinghoff,poul}.
Let $n(x,t)=-\partial_x N(x,t)$ denote the number density of clusters of size $x$ at time $t$.
Smoluchowski's fragmentation-coagulation model describes the dynamics of
clusters subject to fragmentation and coagulation processes. If $K(x,x')$
and $F(x,x')$ represent, respectively, the rate of
coagulating two clusters of size $x$ and $x'$ into a cluster of
size $x+x'$ and the rate of fragmenting a cluster of size $x+x'$ into
two clusters of size $x$ and $x'$, the model takes the form
\begin{eqnarray}
\partial_t n(x,t) &=& 
-n(x,t)\int_0^x F(x-x',x') dx'\nonumber\\
&+&2\int_x^L F(x,x')n(x+x',t)dx'\nonumber\\
&+& \frac{1}{2}\int_0^x K(x-x',x')n(x-x',t)n(x',t)dx'\nonumber\\
&-& n(x,t)\int_0^L K(x,x')n(x',t)dx', 
\label{ML}
\end{eqnarray}
where $L$ represents the largest possible cluster size.
Using dimensionless quantities,  Eq. (\ref{dimless}), 
our model corresponds to the choice, $L=1$, $F(x,x')=1$ and 
$K(x,x')=b \cdot (x \cdot x')$. The 
general form of the merging-fragmentation process then reads
\begin{eqnarray}
\partial_t n(x,t) &=& -xn(x,t)+2N(x,t) \nonumber \\   
&+& \frac{b}{2}\int_0^x x'(x-x')n(x',t)n(x-x',t) dx'\nonumber\\
&-&  b x n(x,t), 
\label{M}
\end{eqnarray}
where $N(x,t)=\int_x^1 n(x',t) dx'$ and the mass normalization,
$\int_0^1 xn(x,t)dx =1$, has been used.
We note that the representation of the dynamics in 
terms of a density function, $n(x,t)$, is only accurate if the typical spacing $\delta x$ between
clusters of size $\sim x$, satisfies $\delta x\ll x$, implying $x\ll 1$. 
In particular,  the assumption of smoothness is 
important for the integral expressions for the fragment gain and loss due to 
merging  to be correct. These terms erroneously include 'self-merging'
processes, $x+x\rightarrow 2x$, as well, thereby  
giving an addition to the actual merging rates which can
only be ignored in the limit $\delta x\ll x$. 
Eq. (\ref{ML}-\ref{M}) are therefore not adequate for 
the dynamics of fragments carrying a large fraction of
the total mass, as the error in the merging rates
become non-negligible here and fail to guarantee total mass conservation.
In fact, Eq. (\ref{ML}), is only mass conserving in the limit $L\rightarrow \infty$,
for which the particular rescaling to dimensionless quantities in
Eq. (\ref{dimless}), can not be done. In the following, we will
tentatively assume the continuous representation to be correct up
to some length scale $l\ll 1$ and seek analytical solutions to
Eq. (\ref{M}) in this regime.

Critical points are generally associated with scale invariant behavior.
For cluster dynamics with merging and fragmentation, scale
invariance has been obtained previously in \cite{vicsek,family,ernst,
source,zheng}, although all under somewhat different circumstances 
i.e. with source terms and thus not in a closed system.  
Therefore, we are searching for stationary scale-invariant
solutions on the form, $n_0(x)\sim x^{\alpha-1}$ 
or $N_0(x)= A x^{\alpha}+C$. Inserting this scaling relation into Eq. (\ref{M}) we obtain
\begin{eqnarray}
0 &=& -Ax^\alpha\left(\alpha+2+b\alpha\right) + 2C \nonumber\\
&+& \frac{b}{2} (\alpha A)^2 x^{(2\alpha+1)} * B(\alpha+1,\alpha+1)
\label{Bal}
\end{eqnarray}
where $B(\alpha+1,\alpha+1)$ is the Beta function
$B(\alpha+1,\alpha+1)=\Gamma(\alpha+1)\Gamma(\alpha+1)/\Gamma(2\alpha+2)$
under the assumption that $\alpha+1 > 0$. Power counting arguments now
easily gives that only  two values for the exponent is possible,
namely $\alpha=-1$ or $\alpha = -1/2$. The value $\alpha=-1$, however
will ruin the assumption $\alpha+1>0$ and is inconsistent with the 
finite mass of the system \cite{comment}. 
Thus we come to the conclusion that {\it there is only one scale invariant solution}
to Eq. (\ref{M}), namely  $n_0(x) = A/2 x^{-3/2}$, 
which becomes an exact solution provided that $\alpha+2+b\alpha=0$
or $b=3$. Also, stationarity requires $C=-3/16A^2\pi$ so 
the total number of clusters larger than $x$ for $x_c(t)\ll x<l$, is simply 
\begin{equation}
N_0(x)=A\left( x^{-1/2} -3/16 A\pi \right).
\label{N}
\end{equation}
The actual value of $A$ depends on how much of the total mass is situated in 
the discrete part of the spectrum given by the few fragments with
$x>l$, $\int_0^l xn_0(x)dx + \sum_{x_i>l} x_i =1$. Setting (incorrectly) $l=1$ implies $A=1$.
To assess the scaling behavior in the large mass limit, we show in
the inset of Fig. \ref{snapshot} the average cumulative distribution
for $x>10^{-3}$ where size fluctuations becomes noticeable. Using
a fitted value of $A\approx 0.9$, the analytical solution, Eq. (\ref{N}) is observed to be 
accurate all the way up to $l\sim 0.1$, as expected from
the condition $l\ll 1$. The value for $l$ corresponds to 
the typical size ($50\%$ fractile) of the third largest fragment, $m_3\approx 0.09$,
effectively meaning that only the two largest fragments belong
to the discrete spectrum. In fact, the distribution $N(x)$ has 
significant gaps around the two largest fragments, e.g. 
 $\delta x=1/2(m_1-m_3)\approx 1/2(0.47-0.09)=0.19$, compared to the 
small differences $\delta x\ll 1$ around all other fragments.

\begin{figure}
\epsfig{file=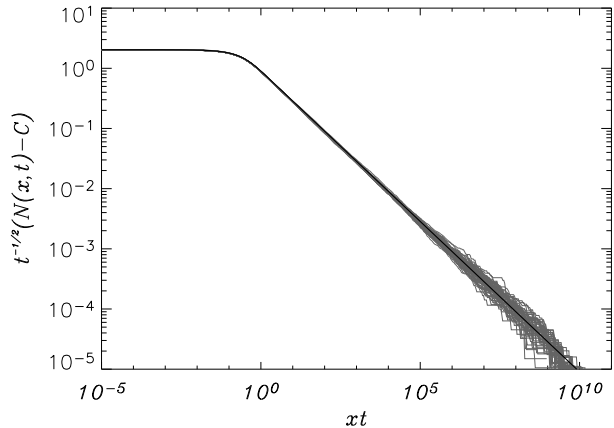,width=.5\textwidth}
\caption{
The rescaled cumulative distribution, $t^{-1/2}(N(x,t)-C)$, 
where $C=-3/16 A^2\pi$, Eq. (\ref{rescale}), as function of $z=xt$ taken at 50 different
times in the interval $10^7\le t \le 2.5\cdot 10^{10}$. 
In black solid line is shown the predicted
functional form $\phi(z)=A\frac{\mbox{erf}(2\sqrt{z})}{\sqrt{z}}$.
}\label{rescale_fig}
\end{figure}

Eq. (\ref{M}) suggests that the full
time dependent distribution, $N(x,t)$ can be expressed
in the form $N(x,t)=t^\gamma\phi(xt)+C$. Indeed, setting $z=xt$ 
one obtains from Eq. (\ref{M}) 
\begin{eqnarray}
\lefteqn{-t^\gamma\left((\gamma+1)\phi'(z)+z\phi''(z)\right)=}  \nonumber \\
 & &  t^\gamma\left(z\phi'(z)+2\phi(z)+bz\phi'(z)\right) + \nonumber \\
 & & C+1/2t^{2\gamma-1} b \int_0^z z'(z-z')\phi'( z)\phi'(z-z')dz',  
\label{rescale0}
\end{eqnarray}
which is a function of $z$ only, provided that either 
$\gamma=1$ or $\gamma=1/2$ and $1/2b\int_0^z z'(z-z')\phi'(
z)\phi'(z-z')dz' \approx -C$. Requiring $N(x,t)=N_0(x)$ for $x\gg x_c(t)$, implies $\phi(z)\sim z^\alpha$ for 
large $z$ and $\gamma=-\alpha=1/2$. Neglecting the two last terms on the 
rhs. of Eq. (\ref{rescale0}), one can obtain the distribution
for all $z$ by solving the linear differential equation,
$$
-3/2\phi'(z)-z\phi''(z) = z\phi'(z)+2\phi(z)+3z\phi'(z), 
$$
where $\gamma=1/2$ and $b=3$ has been inserted. The solution
yields $\phi(z)=a_1z^{-1/2}+a_2\psi(z)$, 
where $\psi(z)=\frac{\mbox{erf}(2\sqrt{z})}{\sqrt{z}}$ 
and $\mbox{erf}(y)$ is the error function,
$\mbox{erf}(y)=\frac{2}{\sqrt{\pi}}\int_0^y e^{-r^2}dr$.
The integration constants, $a_1$ and $a_2$, can be found
from the known behavior of $\phi(z)$ in the limits $z\rightarrow
\infty$ and $z\rightarrow 0$. Requiring $\phi(z)\rightarrow \mbox{const}$
for $z\rightarrow 0$ implies $a_1=0$ and since $\psi(z)\rightarrow z^{-1/2}$ 
for large $z$, one obtains $a_2=A$. For this solution the convolution 
term in Eq. (\ref{rescale0}) is bounded, $0\le \int_0^z
z'(z-z')\phi'(z)\phi'(z-z')dz\le B(1/2,1/2)$ and has a weak
$z$-dependence which is negligible at larger times. 
The final form of $N(x,t)$ then reads
\begin{equation}
N(x,t)=t^{1/2}\phi(xt)- 3/16 A^2\pi, ~~~~\phi(z)=A\frac{\mbox{erf}(2\sqrt{z})}{\sqrt{z}}.
\label{rescale}
\end{equation}
In Figure \ref{rescale_fig}, we show $(N(x,t)+3/16 A^2\pi)t^{-1/2}$, as function
of $z=xt$ for 50 distributions taken in the time interval $10^7\le t
\le 2.5\cdot 10^{10}$. The thick solid line is the predicted functional form,
$\phi(z)$ which agrees perfectly with data. The full time dependent
solution, Eq. (\ref{rescale}), of the combined process
of fragmentation and coagulation is the main result of this Letter. The spread around
$\phi(xt)$ for large values of $xt$ is caused by noticeable
fluctuations of the largest fragments of the distribution. Eq. (\ref{rescale})
also shows that the  characteristic cross-over scale, $x_c(t)$ 
(above which the the system is stationary and scale free),
goes with time as $x_c(t)\sim t^{-1}$. Conversely, had we introduced a smallest length scale,
$\delta$, in the system, we would expect convergence to a full
stationary distribution after time $t\sim \delta^{-1}$, 
displaying scale invariance in the region $\delta \ll x\ll 1$. 

Let us now comment on conserved versus non-conserved
merging/fragmentation processes. Imagine a simulation where clusters,
when becoming too large, above the scale $\tilde l$, are disregarded and 
the total mass is not conserved. From these kind of simulations 
we have observed a transient scale invariant cluster size 
distribution, $n(x)\sim x^{-2}$, consistent with simple 
power counting of Eq. (\ref{M}) (corresponding to $\alpha=-1$). 
However, the total mass will eventually become smaller than $\tilde l$, after which the 
dynamics again is mass conserving leading to the $n(x)\sim x^{-3/2}$
distribution. The $-2$ versus $-3/2$ solutions 
might be relevant in discussions of data from a variety of systems
depending on the degree to which they are mass conserving. 
In the case of American companies, the exponent of the
scale invariant distribution is
closer to $-2$ than $-3/2$ \cite{company}, which in the
light of our model  indicates that
the total wealth of companies is not conserved.

In this letter we have presented a general formalism for scale
invariant and critical behavior in systems where merging balances
fragmentation. We have proposed a dynamical model where the 
merging/fragmentation processes drive the system into a critical state.
From an analytical solution of Smoluchowski's equation we show that 
this scale invariant state represents a unique part of the cluster distribution.
We suggest that the proposed principle could be the dynamical
background behind many scale free systems of society and nature.

We are grateful to S. Manrubia, S. Maslov and K. Sneppen for useful discussions.


\begin{thebibliography}{10}
\bibitem{comment}
To be precise, $\alpha=-1$ implies that
the total mass in the scale-invariant region, $x_c(t)<x<l$,
becomes $\int_{x_c(t)}^l xn_0(x) dx \sim \log(l/x_c(t))\rightarrow
\infty$ as the lower scale, $x_c(t)$, tends to zero at large times.
\end{thebibliography}
\end{document}